\documentclass[lettersize,journal]{IEEEtran}
\usepackage{amsmath,amsfonts, amssymb}
\usepackage{array}
\usepackage[caption=false,font=normalsize,labelfont=sf,textfont=sf]{subfig}
\usepackage{textcomp}
\usepackage{stfloats}
\usepackage{url}
\usepackage{verbatim}
\usepackage{graphicx}
\usepackage{tabularx}
\usepackage{booktabs}
\usepackage{algorithm}
\usepackage[noend]{algpseudocode}
\usepackage{multirow}
\usepackage{makecell}
\usepackage{lipsum}
\usepackage{scalerel}
\usepackage{comment}
\usepackage{enumitem}
\usepackage{threeparttable}
\setlist[itemize]{noitemsep, nolistsep}
\setlist[enumerate]{noitemsep, nolistsep}
\usepackage{color}

\usepackage{balance}

\usepackage{pifont}

\usepackage[utf8]{inputenc}
\usepackage{kotex}

\newcommand{\Numex}{$N_{ex}$\xspace}

\newcommand{\topk}{top-$k$\xspace}

\definecolor{darkorange}{RGB}{255,97,3}

\usepackage{xspace}

\begin{document}

\title{SSD Offloading for LLM Mixture-of-Experts Weights Considered Harmful in Energy Efficiency}

\author{Kwanhee~Kyung,~Sungmin~Yun,~and~Jung~Ho~Ahn,~\IEEEmembership{Senior Member, IEEE}

\thanks{This work was partly supported by the Mobile eXperience (MX) Business, Samsung Electronics Co., Ltd and the MSIT (Ministry of Science, ICT), Korea, under the Global Scholars Invitation Program (RS-2024-00456287) supervised by the IITP (Institute for Information \& Communications Technology Planning \& Evaluation).}
\thanks{Kwanhee~Kyung, Sungmin~Yun and Jung Ho Ahn are with Seoul National University, Seoul 08826, South Korea. E-mail: \{kwanhee5, sungmin.yun, gajh\}@snu.ac.kr. Jung Ho Ahn is the corresponding author.}
}


\maketitle
\begin{abstract}
Large Language Models (LLMs) applying Mixture-of-Experts (MoE) scale to trillions of parameters but require vast memory, motivating a line of research to offload expert weights from fast-but-small DRAM (HBM) to denser Flash SSDs. While SSDs provide cost-effective capacity, their read energy per bit is substantially higher than that of DRAM. This paper quantitatively analyzes the energy implications of offloading MoE expert weights to SSDs during the critical decode stage of LLM inference. Our analysis, comparing SSD, CPU memory (DDR), and HBM storage scenarios for models like DeepSeek-R1, reveals that offloading MoE weights to current SSDs drastically increases per-token-generation energy consumption (e.g., by up to $\sim$12$\times$ compared to the HBM baseline), dominating the total inference energy budget. Although techniques like prefetching effectively hide access latency, they cannot mitigate this fundamental energy penalty. We further explore future technological scaling, finding that the inherent sparsity of MoE models could potentially make SSDs energy-viable \emph{if} Flash read energy improves significantly, roughly by an order of magnitude.

\end{abstract}


\vspace{-0.25in}
\section{Introduction}
\label{sec:introduction}

Large Language Models (LLMs) have advanced significantly by populating more pretrained weight parameters to enhance inference accuracy~\cite{2020-arxiv-scaling-law-nlp}.
This trend, however, amplifies both the memory capacity needed for weight and the data volume loaded from memory per-token-generation.
Thus, achieving high energy efficiency alongside high memory bandwidth has become crucial for sustainable, responsive LLM inference services.
Typical data center systems (e.g., the NVIDIA DGX H100) employ a memory hierarchy including device memory (e.g., HBM), CPU memory (e.g., DDR), and NAND Flash-based storage (SSD).
While Flash SSDs offer a lower capacity cost per bit than DRAM, they inherently possess higher read energy consumption and lower data read bandwidth~\cite{2024-arxiv-relu}.

Mixture-of-Experts (MoE) architectures, used in prominent models like DeepSeek-R1, Mixtral, and Llama 4, have become popular for scaling model parameters while controlling computational costs during training.
In MoE models, expert weights constitute the vast majority of parameters (e.g., 96.1\% in DeepSeek-R1).
However, during inference, only a small subset of the experts is activated per token (3.5\% in DeepSeek-R1), leading to a sparse access pattern.
Techniques such as MoE-prefetching~\cite{2024-isca-pregated} exploit this by offloading the large set of expert weights to SSD or CPU memory, aiming to hide the data transfer latency by overlapping it with computation on the active experts.
In particular, while latency may be hidden, the \emph{energy} consumed during these data transfers is not reduced, especially when accessing energy-hungry SSDs.
Since sparsity is most pronounced during the auto-regressive decode stage (where one token is generated at a time), this study focuses on analyzing the energy impact within this stage.

\begin{figure}[!tb]
  \center
  \includegraphics[width=0.8\columnwidth]{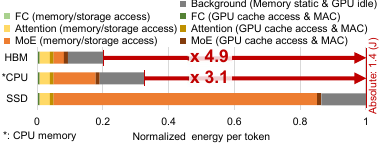}
  \vspace{-0.1in}
  \caption{Normalized per-token energy consumption during the first decode stage after the prefill stage for DeepSeek-R1, comparing MoE weight storage on HBM (baseline), CPU memory (DDR), and SSD. Batch size is 1024 and each request has 1024 input prompt tokens.
  }
  \label{fig:intro}
  \vspace{-0.15in}
\end{figure}

Our analysis shows that offloading MoE weights to SSDs significantly increases energy use.
Inference on DeepSeek-R1 with expert weights stored on SSD consumes approximately 4.9$\times$ more energy per token compared to keeping them in HBM, and 3.1$\times$ more than offloading to CPU memory (Fig.~\ref{fig:intro}).
The energy for accessing weights on the SSD accounts for a staggering 80\% of the total per-token energy in this scenario, underscoring the severity of the energy penalty.

Despite this current drawback, if future Flash technologies achieve better energy efficiency (even if still higher than DRAM), the combination of MoE sparsity and SSD capacity might offer a path to running highly accurate, large models more efficiently than smaller dense models.
Thus, this work provides a quantitative analysis of layer-wise energy consumption and execution time in MoE models.
We compare energy and performance based on the weight storage location, demonstrating that the high energy cost of SSD access is not concealed by latency-hiding techniques.
Further, we project the impact of potential Flash energy improvements on the feasibility of SSD offloading.

\vspace{-0.025in}
\section{Energy Comparison: DRAM vs. Flash}
\label{sec:motivation}
\vspace{-0.025in}
NAND Flash memory's read operation inherently consumes more energy per bit compared to DRAM.
This difference stems from the underlying technologies: Flash requires applying relatively high voltages and using complex mechanisms to sense a cell's threshold voltage, whereas DRAM relies on low-voltage differential sensing of capacitor charge~\cite{ddr5,2010-date-flash}.
This higher read energy characteristic of Flash significantly contributes to the increased energy consumption observed when offloading LLM weights to SSDs.

\begin{figure}[!tb]
  \center
  \includegraphics[width=0.6\columnwidth]{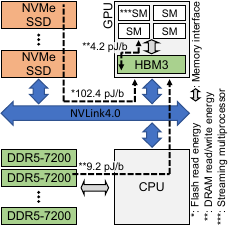}
  \caption{
  The energy consumption during read/write operations on data stored in device memory, CPU memory, and SSD.
  }
  \label{fig:energy_component}
  \vspace{-0.15in}
\end{figure}

Fig.~\ref{fig:energy_component} compares the energy components (in pJ/b) for accessing data stored in device memory (HBM3), CPU memory (DDR5-7200), and an NVMe SSD within a representative server system (specifications detailed in Section~\ref{sec:setup}).
CPU, GPU, and SSD are interconnected via NVLink, similar to NVIDIA's Grace Hopper Superchip~\cite{gh200}.
The values account for both the energy consumed within the memory/storage chips and that consumed in the external I/O paths.
Specifically, internal Flash read energy is derived from dividing energy per read operation by page size, based on \cite{2010-date-flash}.
Internal DRAM energy is calculated using industry datasheets~\cite{ddr5,ddr-power}, and prior work~\cite{2023-vlsi-hbm3}, aggregating data transfer and activation/precharge energy.
External I/O energy includes NVLink energy and memory interface energy (calculated from datasheets or scaled based on prior work~\cite{2017-micro-fgdram}).

In a baseline where LLM weights reside entirely in the GPU's HBM, accessing those consumes only the HBM read energy, estimated at 4.2pJ/b.
By contrast, accessing weights offloaded to an SSD involves a sequence of operations, even when utilizing technologies like GPUDirect Storage~\cite{gds} that bypass intermediate copies to CPU memory.
It includes the Flash read operation itself, writing the data into HBM, and finally reading it from HBM for computation.
The total energy becomes 102.4pJ/b (Flash read) + 4.2pJ/b (HBM write) + 4.2pJ/b (HBM read) = 110.8pJ/b, representing a substantial $\sim$26-fold increase in weight access energy compared to the HBM-only baseline.

\section{LLM, MoE, and MoE-Prefetching}
\label{sec:background}

\begin{figure}[!tb]
  \center
  \includegraphics[width=0.87\columnwidth]{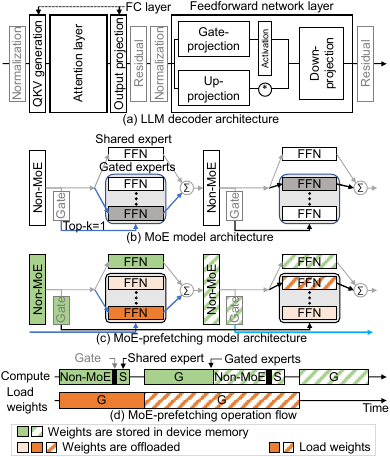}
  \caption{
  (a) High-level structure of an LLM decoder layer. (b) MoE layer showing the gate and multiple experts. (c) MoE model with gated expert weights offloaded. (d) MoE-prefetching operation overlapping computation with background weight loading.
  }
  \label{fig:priorwork}
  \vspace{-0.15in}
\end{figure}

\noindent \textbf{LLM:}
Modern LLMs predominantly consist of stacked decoder layers.
Each decoder typically includes: 1) Attention layers that compute relationships between tokens using query, key, and value projections; 2) Feedforward Network (FFN) layers consisting of gate-projection, up-projection, activation, and down-projection linear transformations; and 3) supporting operations such as layer normalization and residual connection (see Fig.~\ref{fig:priorwork}(a)).
LLM inference proceeds in two main stages: a prefill stage that processes the input prompt tokens ($L_{in}$) in parallel, and a decode stage that auto-regressively generates output tokens ($L_{out}$) one by one, using the previously generated token as input.

\noindent \textbf{MoE:}
MoE architectures (e.g., Mixtral, DeepSeek-R1, and Llama 4) replace most FFN layers with MoE layers.
An MoE layer contains multiple parallel FFNs, termed `experts' ($N_{ex}$).
For each input token, a lightweight `gate' network dynamically selects a small subset (top-$k$) of these experts (`gated experts').
The token is processed only by the selected experts (and few `shared experts' processed by all tokens), and their outputs are combined via a weighted sum determined by the gate values (see Fig.~\ref{fig:priorwork}(b)).
This token-dependent expert activation leads to sparsity: only a fraction of the total MoE weights are accessed per token (e.g., $\sim$25.0\%, $\sim$3.5\%, and $\sim$1.6\% for Mixtral 8x7B, DeepSeek-R1, and Llama 4 Maverick, see Table~\ref{tab:model_configuration}).
However, when processing requests in batches, more experts are activated as different tokens may select different experts.

\noindent \textbf{MoE-prefetching:}
To manage the large memory footprint of MoE weights, prefetching techniques~\cite{2024-isca-pregated} offload a (typically large) set of gated expert weights to lower tiers of the memory hierarchy, such as CPU memory or SSDs (see Fig.~\ref{fig:priorwork}(c)).
Those involve fine-tuning the gate layer of the i-th decoder ($dec_i$) to predict which experts will be needed in the subsequent decoder, $dec_{i+1}$.
While $dec_i$ and the initial layers of $dec_{i+1}$ are executing, the system concurrently fetches the predicted expert weights for $dec_{i+1}$ from the offload storage (CPU memory or SSD) into the GPU's device memory (HBM).
This overlap aims to hide the data transfer latency (see Fig.~\ref{fig:priorwork}(d)).

\section{Experimental setup}
\label{sec:setup}

\begin{table}[tb!]
  \centering
  \caption{Inference System and Model Configuration}
  \vspace{-0.05in}
  \label{tab:model_configuration}
  \resizebox{0.90\columnwidth}{!}{
    \Huge
    \begin{tabular}{r||c|c|c|c|c|c|c}
      \Xhline{2pt}
       Model        & Param. & \# Node & \makecell{\# GPU \\ per node} & \makecell{\# FFN \\ decoder} & \makecell{\# MoE \\ decoder} & \Numex               & \topk \\
      \midrule
       Mixtral      & 47B  & 1 & 4 & –  & 32      & 8 (*G)               & 2 \\
       DeepSeek-R1  & 671B  & 4 & 8 & 3  & 57        & 256 (G) + 1 (**S)   & 8 \\
       \midrule
       Llama 4          & 400B & 1 & 8  & 24  & 24     & 128 (G) + 1 (S)  & 1 \\
       Llama 3.3       & 70B  & 1 & 8  & 80  &    –    & –                    & – \\
      \Xhline{2pt}
      \multicolumn{8}{c}{*G: Gated expert \quad **S: Shared expert}
    \end{tabular}
  }
  \vspace{-0.05in}
\end{table}

\noindent \textbf{Inference system modeling:}
We used a modeling tool\footnote{It is located at \url{https://github.com/scale-snu/SSD-offloading}}---modified from the simulator introduced in \cite{2024-micro-duplex}---which incorporates CPU memory/SSD read bandwidth and the resulting reduction in HBM read bandwidth when MoE prefetching operations overlap.
The compute node mirrors a DGX H100, featuring up to eight NVIDIA H100 SXM5 80GB GPUs interconnected with NVLink4.0 (450GB/s per direction).
In multi-node setups, nodes are connected via InfiniBand using eight ports, each delivering 50GB/s per direction.
Each GPU is associated with 256GB of CPU memory (DDR5-7200) and NVMe SSD (see Fig.~\ref{fig:energy_component}).
The aggregated read bandwidth from CPU memory and SSD to the GPU matches the unidirectional bandwidth of the NVLink interconnect, leveraging technologies such as DMA and GPUDirect Storage, respectively.

\noindent \textbf{Models and configurations:} All experiments employ the BF16 data type for model weights and activations.
Section~\ref{subsec:amplification} focuses on the MoE models of Mixtral 8x7B (referred to as Mixtral) and DeepSeek-R1.
The baseline is configured with sufficient device memory to hold the entire model.
Section~\ref{subsec:scaling} compares the Llama 4 Maverick MoE model (referred to as Llama 4) with the Llama 3.3 non-MoE model.
We simulated a scenario where only the smaller Llama 3.3 model fits entirely within device memory, requiring offloading for Llama 4.
In the models used for the experiments, shared experts and non-MoE FFN layers are both categorized alongside other Fully-Connected (FC) layers.
Table~\ref{tab:model_configuration} summarizes the inference system and model configuration.

\noindent \textbf{Parallelism and communication:}
We adopt parallelism strategies from~\cite{2024-micro-duplex}.
The MoE layers utilize expert parallelism, assigning distinct experts to different GPUs.
The non-MoE FFN/FC layers and shared experts employ tensor parallelism within nodes and data parallelism across nodes.
To synchronize data across GPUs, the output projection, the non-MoE FFN/MoE layers, and the shared experts in MoE layers perform an intra-node AllReduce after their operations. 
Further, the MoE layers perform dispatch and combine to compute with gated experts distributed across different GPUs.

\noindent \textbf{Simulation details:}
We used synthesized datasets, assuming that, for each decoder, token-to-expert assignments during simulation follow a uniform distribution.
For all simulated requests, we focused on the first decode stage, after the prefill stage with an $L_{in}$ of 1024 tokens.

\section{Evaluation}
\label{sec:evaluation}
\vspace{-0.05in}

First, we analyze the energy and latency impacts of offloading MoE weights to SSDs compared to HBM and CPU memory, revealing that the system energy consumption significantly deteriorates when weight offloading to SSDs.
Second, we investigate the potential for energy-efficient SSD utilization under future Flash technology improvements by leveraging MoE sparsity.

\begin{figure}[!tb]
  \center
\includegraphics[width=0.99\columnwidth]{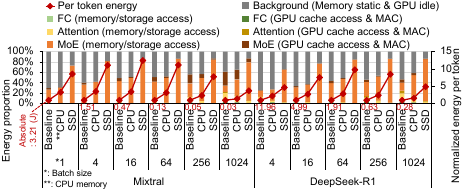}
  \vspace{-0.25in}
  \caption{
  Energy breakdown and normalized per-token energy. The CPU memory and SSD are interconnected to the GPU via NVLink5.0.
  }
  \label{fig:server_energy}
  \vspace{-0.1in}
\end{figure}

\subsection{Impacts of Flash Offloading on Energy and Latency}
\label{subsec:amplification}

We evaluated Mixtral and DeepSeek-R1, comparing three scenarios for gated expert weights: baseline (stored in HBM), offloaded to CPU memory, and offloaded to SSD.

\noindent \textbf{Energy consumption:}
We divided the energy breakdown into three main components: memory/storage access; GPU compute, defined as the sum of cache access and MAC operations; and system-wide background energy, which combines memory static power (57W per 256GB of CPU memory~\cite{ddr5,ddr-power}) and GPU idle power (70W per GPU) consumption. SSD idle power is negligible~\cite{ssd-idle-power} and thus excluded.
As shown in Fig.~\ref{fig:server_energy}, \emph{offloading gated expert weights to SSD leads to a dramatic increase in the energy consumed per generated token during the decode stage.}
Compared to the baseline, SSD offloading results in 3.8$\times$ to 12.5$\times$ higher per-token energy for Mixtral and 4.7$\times$ to 9.8$\times$ higher energy for DeepSeek-R1, varying with batch size.
Even compared to offloading to CPU memory, SSD offloading is significantly worse, costing 2.1$\times$ to 3.6$\times$ more energy.
This substantial energy penalty directly attributes to the high energy cost of reading from Flash memory.

As the batch size increases, the total energy for loading the activated MoE experts per batch tends to saturate once most or all unique experts ($N_{ex}$) are required across the batch (observed around batch size 16 for Mixtral and 64 for DeepSeek-R1 in our tests).
This means that the per-token MoE memory/storage access energy decreases with larger batches due to amortization.
By contrast, GPU compute energy for FC, Attention, and MoE layers increases with batch size due to the increasing computational workload.
Despite the amortization effect, the amplified energy for accessing weights on the SSD remains the dominant factor at large batch sizes (1024): it accounts for over 73\% (Mixtral) and 80\% (DeepSeek-R1) of the total energy with SSD offloading, whereas in the baseline, the energy to access MoE weights stored in HBM drops to only 11\% and 15\%, respectively.
Importantly, latency-hiding via prefetching does not alter these underlying energy consumption figures.

\begin{figure}[!tb]
  \center
\includegraphics[width=0.9\columnwidth]{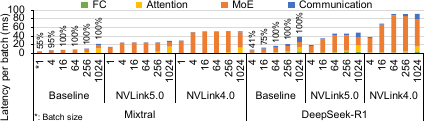}
  \vspace{-0.1in}
  \caption{
  Latency breakdown for generating a single token in a batch.
  The percentile above the graph is the ratio of the sum of the maximum active gated expert counts on a single GPU per decoder, to the equivalent total sum when all experts are activated.
  }
  \label{fig:server_latency}
  \vspace{-0.1in}
\end{figure}

\noindent \textbf{Token generation latency:}
While energetically costly, SSD offloading's impact on latency can be effectively managed.
Modern high-bandwidth interconnects like NVLink reduce the raw data transfer time from SSDs and CPU memory.
Further, prefetching techniques effectively hide much of this remaining transfer latency by overlapping it with the execution of other layers.
Fig.~\ref{fig:server_latency} shows that with efficient prefetching and a fast interconnect like NVLink5.0 (doubling NVLink4.0 bandwidth), the end-to-end token generation latency per batch when offloading can be comparable to the baseline configuration (with the latency penalty reduced to as low as 1.32$\times$ for Mixtral and 1.25$\times$ for DeepSeek-R1).
This indicates a trade-off: sacrificing energy efficiency for manageable latency and the ability to use larger models than fit in device memory.

\vspace{-0.1in}
\subsection{Opportunities from Flash Energy Scaling}
\label{subsec:scaling}

\begin{figure}[!tb]
  \center
\includegraphics[width=0.9\columnwidth]{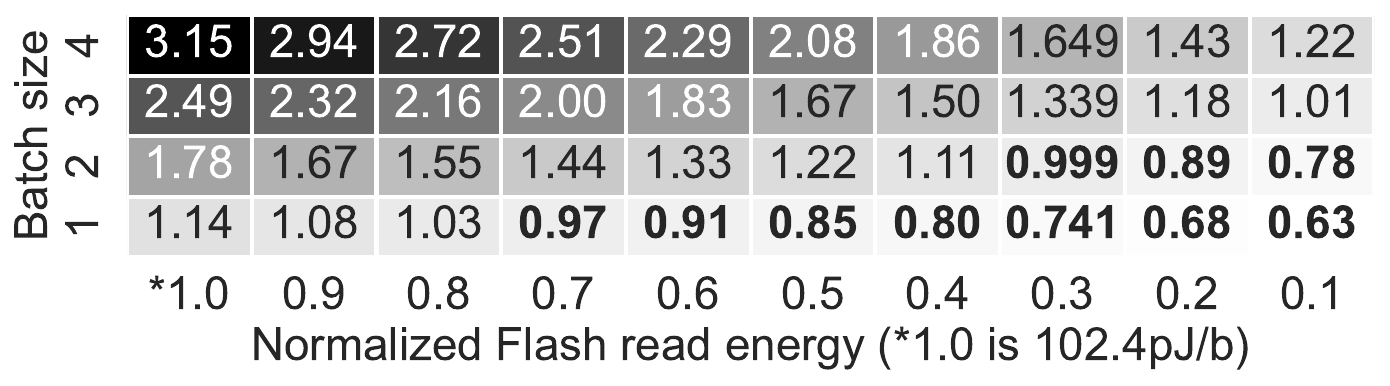}
  \vspace{-0.15in}
  \caption{
  Per-token energy consumption ratio of Llama 4 to Llama 3.3 with respect to batch size and Flash read energy. The SSD is interconnected to the GPU via NVLink5.0.
  }
  \label{fig:heatmap}
  \vspace{-0.1in}
\end{figure}

Section~\ref{subsec:amplification} showcased the high energy cost of SSD offloading with current technology.
Here, we explore whether future improvements in Flash read energy efficiency could alter this conclusion, potentially making SSDs attractive for large MoE models.
We compare the energy efficiency of inferring a large MoE model, Llama 4 (400B), requiring SSD offloading, against a smaller but dense non-MoE model, Llama 3.3 (70B), assumed to fit entirely in HBM.
Larger MoE models like Llama 4 are expected to provide higher inference accuracy than smaller non-MoE models like Llama 3.3.

The key idea is \emph{leveraging MoE sparsity}: if the number of parameters accessed per token in the large MoE model (due to sparsity) is sufficiently smaller than the total parameters of the dense model, the MoE model might achieve lower per-token energy despite using the less efficient SSD, provided the SSD's energy penalty is reduced.
For Llama 4 at batch size one, only 17B parameters (4.3\% of the total) are activated per token, with 17.8\% of these activated weights coming from the offloaded gated experts on SSD.
By contrast, Llama 3.3 always accesses its full 70B parameters.
As batch size increases for Llama 4, more experts are activated, eventually accessing all 400B parameters when the batch is large enough to activate all $N_{ex}$ experts.

Fig.~\ref{fig:heatmap} plots the per-token energy ratio of Llama 4 to Llama 3.3 as a function of the Flash read energy scaling factor, in a system with only GPU and SSD.
Llama 4 on SSD becomes more energy-efficient than Llama 3.3 on HBM (ratio $< 1$) only when the Flash read energy is significantly reduced---specifically, to approximately 1/10th of its current value ($\sim$10pJ/b).
This ``SSD winning case'' is most prominent at low batch sizes (batch size $< 3$ in this experiment), where MoE sparsity is highest,
showcasing that substantial technological advancements in Flash energy efficiency are required before SSD offloading becomes energetically competitive with HBM-based inference, even when exploiting MoE sparsity.

\vspace{-0.05in}
\section{Discussion}
\label{sec:discussion}

Our findings highlight a critical trade-off: current SSDs impose high energy costs for MoE weight offloading (Section~\ref{subsec:amplification}), but future Flash advancements could shift this balance.
Analysis (Section~\ref{subsec:scaling}) indicates large MoE models on SSDs might achieve better energy efficiency than smaller dense models, provided Flash read energy improves by $\sim$10$\times$ (to $\sim$10pJ/b) and MoE sparsity is effectively exploited.

This scenario is particularly relevant to mobile/edge platforms,
which typically operate at small batch sizes, maximizing MoE sparsity, face strict battery constraints, and possess analogous LPDDR/UFS memory hierarchies.
Thus, if Flash reaches the $\sim$10pJ/b read energy target, mobile represents a prime domain to potentially benefit from large, accurate MoE models using energy-efficient Flash offloading.
Realizing this requires continuous evolution in Flash focused on energy reduction and system co-design for sparse data management.

\section{Conclusion}

Offloading MoE weights to contemporary Flash SSDs results in a surge in LLM inference energy consumption due to high Flash read energy.
While latency can often be hidden, this energy penalty makes SSD offloading energetically unfavorable today.
However, significant future improvements in Flash read energy efficiency (approximately 10$\times$ reduction) could, combined with the inherent sparsity of MoE models at low batch sizes, potentially make SSDs an energy-viable option for deploying very large, high-accuracy LLMs.

\bibliographystyle{IEEEtran}
\bibliography{ref}

\end{document}